\documentclass{emulateapj}

\usepackage{graphicx}
\usepackage{epstopdf}

\newcommand{\lx}{erg~s$^{-1}$}

\slugcomment{To appear in ApJ, July 1st, 2014 issue.  Vol. 789, p. 64}

\shorttitle{IC10 X-ray Transient}
\shortauthors{Laycock et al.}

\begin{document}


\title{A Transient Supergiant X-ray Binary in IC10. An Extragalactic SFXT?}


\author{Silas Laycock, Rigel Cappallo, Kathleen Oram, and Andrew Balchunas}
\affil{Dept of Physics and Applied Physics, Olney Science Center, University of Massachusetts Lowell, MA, 01854 }



\begin{abstract} 
We report the discovery of a large amplitude (factor of $\sim$100) X-ray transient (IC 10 X-2, CXOU J002020.99+591758.6) in the nearby dwarf starburst galaxy IC10 during our Chandra monitoring project. Based on the X-ray timing and spectral properties, and an optical counterpart observed with Gemini, the system is a high mass X-ray binary (HMXB) consisting of a luminous blue supergiant and a neutron star (NS). The highest measured luminosity of the source  was 1.8$\times$10$^{37}$ \lx during an outburst in 2003. Observations before, during and after a second outburst in 2010 constrain the outburst duration to be less than 3 months (with no lower limit). The X-ray spectrum is a hard powerlaw ($\Gamma$=0.3) with fitted column density ($N_H$=6.3$\times$10$^{21}$ atom cm$^{-2}$) consistent with the established absorption to sources in IC10.  The optical spectrum shows hydrogen Balmer lines strongly in emission, at the correct blueshift (-340 km/s) for IC10. The NIII triplet emission feature is seen, accompanied by He II [4686] weakly in emission. Together these features classify the star as a luminous blue supergiant of the OBN subclass, characterized by enhanced nitrogen abundance.  Emission lines of HeI are seen, at similar strength to H$\beta$.  A complex of FeII permitted and forbidden emission lines are seen, as in B[e] stars. The system closely resembles galactic supergiant fast X-ray transients (SFXTs), in terms of its hard spectrum, variability amplitude and blue supergiant primary. 

\end {abstract}


\section{Introduction}

The dwarf starburst galaxy IC10 provides a unique setting to study X-ray binaries containing youngest and most massive stars. According to  \cite{massey2007a} the IC 10 starburst peaked $\sim$7$\times 10^{6}$yrs ago and the galaxy is filled with diffuse HII emission, HII regions, wind-bubbles, and an unusually high space-density of Wolf-Rayet stars \citep{bauerbrandt2004}. 
Young stellar populations such as in IC 10, host massive stars that are rapidly approaching the end of their main sequence lifetimes. Binary systems containing recent post-supernova relics are known to form in such environments, where they can provide an important probe of massive-star evolution. IC 10 contains $\sim$100 X-ray point sources, whose combined spectrum is indicative of a population of High Mass X-ray Binaries \citep{wang2005}.

High mass X-ray binaries (HMXBs: see e.g. \citealt{liu2005, reig2011} for reviews of the field) are binary stars containing a neutron star  (or black hole) and a massive star as the mass-donor. In the majority (upward of 80\%) of known systems in the Milky Way and Magellanic Clouds the primary is a Be star \citep{mcbride2008}. In the remaining systems it is an O or B supergiant, whose wind provides the fuel for accretion-powered X-ray emission. SG-HMXBs emit X-rays continuously at a $10^{35}-10^{36}$ \lx, with eclipses being relatively common as a consequence of the small orbital separations on the order of a few stellar radii. Tidal interactions rapidly circularize the orbits leading to little variation in the wind density and velocity encountered by the compact object.  Be-HMXBs are transient due to a complex interaction between the wide (many A.U.) elliptical orbit and the circumstellar disk surrounding the Be star \citep{okazaki2001}. Frequent outbursts reach $> 10^{37}$ \lx for days at a time and generally recur with the orbital period of the system, taking place at, or close to the time of periastron passage. Much rarer ``Giant" outbursts last for weeks to months, exceed 10$^{37}$ \lx and generally begin around periastron but otherwise show little correlation with orbital parameters. A small number of BH-HMXBs are known, and all have supergiant primaries, in line with population modeling predictions \citep{belczynski2009}.

In recent years a small subset of HMXBs have been discovered that do not fit the picture described above. These are the Supergiant Fast X-ray Transients (SFXTs) (\citealt{negueruela2005,sguera2005}), which share the following properties: Very short duration X-ray flares of $\sim$hours with $L_X$ reaching $10^{37}$ \lx, variability within these flares, hard X-ray spectra, supergiant companions, location in young heavily dust-obscured star-forming regions in the plane of the Milky Way (all but 2 are in the Sagittarius-Scutum Arm at RA=16$^h$-19$^h$).  In between outbursts SFXTs continue to accrete matter at a much lower level, with an average luminosity of 10$^{33}$- 10$^{34}$\lx, occasionally reaching a true quiescent state at $10^{32}$\lx  \citep{Sidoli2008}. It is difficult to explain the abrupt luminous flares, since this requires that either the mass transfer rate from the donor changes by a factors of 10$^2$ - 10$^3$ on short time scales (for example by clumpy stellar wind) or that the accretion onto the NS is somehow gated, perhaps by a magnetic process \citep{Sidoli2009}.
A recent survey by \cite{Romano2014} lists 12 confirmed SFXTs; their orbital periods ranging from 3d (IGR J16479-4514: \citealt{Jain2009})  to 164d (IGR J11215-5952: \citealt{Romano2009}) with most at the shorter end of this range (Half are below 10d  e.g. AX J1845:   \citealt{Goosens2013}). 

X-ray surveys of nearby galaxies have revealed populations of HMXBs in all star-forming galaxies, with their number scaling with star-formation rate \citep{GGS}. There is growing evidence that HMXBs with B-type primaries (the most common type) are strongly associated with stellar populations aged 40-80 million years (\citealt{antoniou2010, williams2013}). The lifetime of an O-supergiant is only few My (depending on the initial mass and chemical composition), on this basis we expect IC10 to contain SG-HMXBs, including SFXTs, which occur in similar regions of the Milky Way.

In this paper we adopt the distance to IC 10, $D_{IC10}$=660 kpc  \citep{wilson1996} which implies a distance modulus $\mu_{IC10}$ = 24.1.
The foreground extinction to sources in IC 10 is $E_{B-V}$=0.85 (e.g. \citealt{Goncalves2008, Sanna2008}) implying $A_V$=3.1

\section{Observations and Data Analysis}
\subsection{Chandra X-ray Data}

We discovered a transient X-ray point-source in a series of 10 Chandra ACIS-S observations of IC 10, spanning 2003-2010 as detailed in Table~\ref{tab:dataset} which include 3 deep observations from the archive, and a year-long monitoring campaign of seven 15 ksec observations at roughly 6 week intervals \citep{Laycock2010}. Data were taken in Very Faint (VF) or Faint (F) mode with the maximum-efficiency 3.2 s CCD frame time. 

\begin{deluxetable}{llllll}
\tablecaption{X-ray Observations of IC10 X-2 \label{tab:dataset}} 
\tablehead{ \colhead{MJD}  &   \colhead{Date} &   \colhead{ObsID}         & \colhead{Offset}     &   \colhead{Exposure} & Mode \\
                                                    &                     &                                      &   \colhead{arcmin} &      \colhead{ksec}   & \colhead{\tablenotemark{a}} }
\startdata
52710.7 & 2003 Mar 12   &  03953 & 1.21     & 28.9  & VF \\
 54041.8 & 2006 Nov 2   &  07082 & 2.48     & 40.1  & F\\
54044.2 & 2006 Nov 5  &  08458 & 2.48     & 40.5   & F\\
55140.7 & 2009 Nov 5 &  11080 & 0.52    & 14.6  & VF\\
55190.2 & 2009 Dec 25  & 11081 & 0.25        & 8.1   & VF \\
 55238.5 & 2010 Feb 11  &11082 & 0.87        & 14.7   & VF\\
 55290.6 & 2010 Apr 4  &11083 & 1.97         & 14.7  & VF\\
55337.8 & 2010 May 21 &11084 &  2.33   & 14.2    & VF\\
55397.5 & 2010 July 20 &11085 & 1.75        & 14.5   & VF\\
55444.6 & 2010 Sept 5  &11086 & 0.78        & 14.7  & VF\\
\enddata
\tablenotetext{a}{Chandra ACIS-S observations in Faint (F) and Very Faint (VF) mode share the same 3.2 s frame-time.}
\end{deluxetable}

The optical diameter of IC 10 is $\sim$7 arcmin, so essentially the whole galaxy fits within the ACIS-S3 chip, where the PSF, sensitivity and spectral resolution are maximized. Chandra's aim-point and roll-angle were varied among the observations,  to best position IC10 within the field of view given scheduling constraints. 

\begin{figure}
\begin{center}
\includegraphics[angle=0,width=8.5cm]{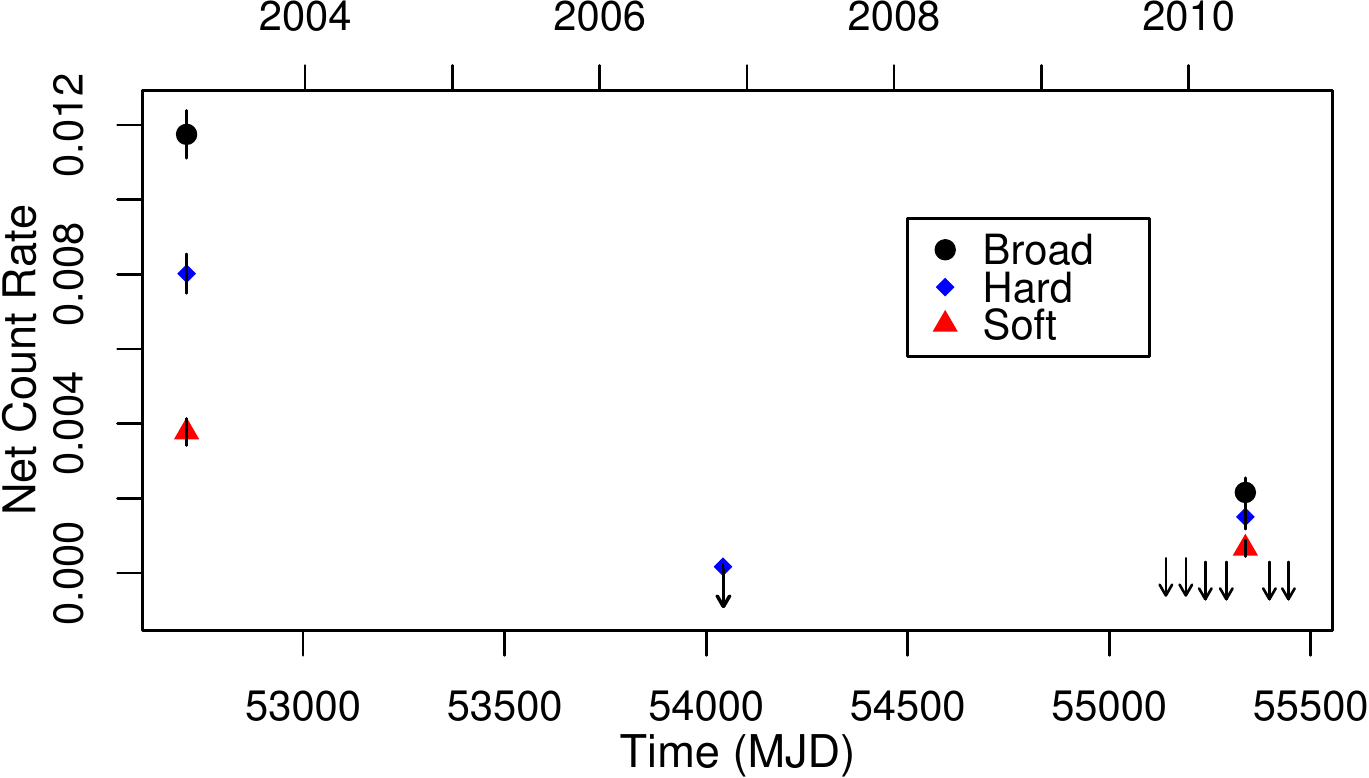}
\caption{{\bf  Chandra 2003-2010 X-ray Lightcurve. Points with statistical error bars show net ACIS-S count rates in Broad (0.3-8 keV), Soft (0.3-2.5 keV) and Hard (2.5-8) energy bands. Upper limits for non-detection in the B-band are shown as arrows. }}
\label{fig:monitor}
\end{center}
\end{figure}

Data reduction was carried out in Ciao 4.3\footnote{http://cxc.harvard.edu/ciao} following the standard processing threads and referring to \cite{hong2005} for consistent selection of energy bands, error circles, and extraction regions. A check of the Ciao release notes indicates that subsequent Ciao versions will not lead to different results. Source detection was accomplished using {\it wavdetect} at wavelet scales of 2,4,8,16 pixels. Exposure maps generated for 1.5 keV were used to correct for sensitivity variations and to mask areas where the exposure is below 10\% of its maximum value. Wavdetect was run on 1x1 binned images in 3 energy ranges: broad-band = 0.3-8 keV, soft-band = 0.3-2.5 keV, and hard-band = 2.5-8 keV. The resulting long-term light curve displayed in Figure~\ref{fig:monitor} is discussed in Section~\ref{xrayresults}.

Astrometric calibration was performed by matching the {\it wavdetect} point-source catalog from the deepest observation (the stacked 2006 data) against the optical catalog of \cite{massey2007a} using the Starbase\footnote{http://hopper.si.edu/wiki/mmti/Starbase} relational database software. We then bore-sighted the source catalogs from each individual Chandra observation onto this calibration using the X-ray coordinates of IC 10 X-1 (which has by far the smallest positional uncertainty owing to its high count-rate).  The final RMS scatter between X-ray and optical coordinates (after rejecting the 3$\sigma$ outliers) is 0.5''.

Subsequent analysis of events, light-curves, spectra and images were carried out using Ciao, {\em R}\footnote{The R-Project for Statistical Computing. www.r-project.org}, XSPEC 12.7\footnote{http://heasarc.nasa.gov/xanadu/xspec} and SAOimage DS9\footnote{http://ds9.si.edu}.

% FIGURE 1       

\subsection{Gemini: Optical Imaging and Spectra}
\begin{figure*}
\begin{center}
\includegraphics[angle=0,width=14cm]{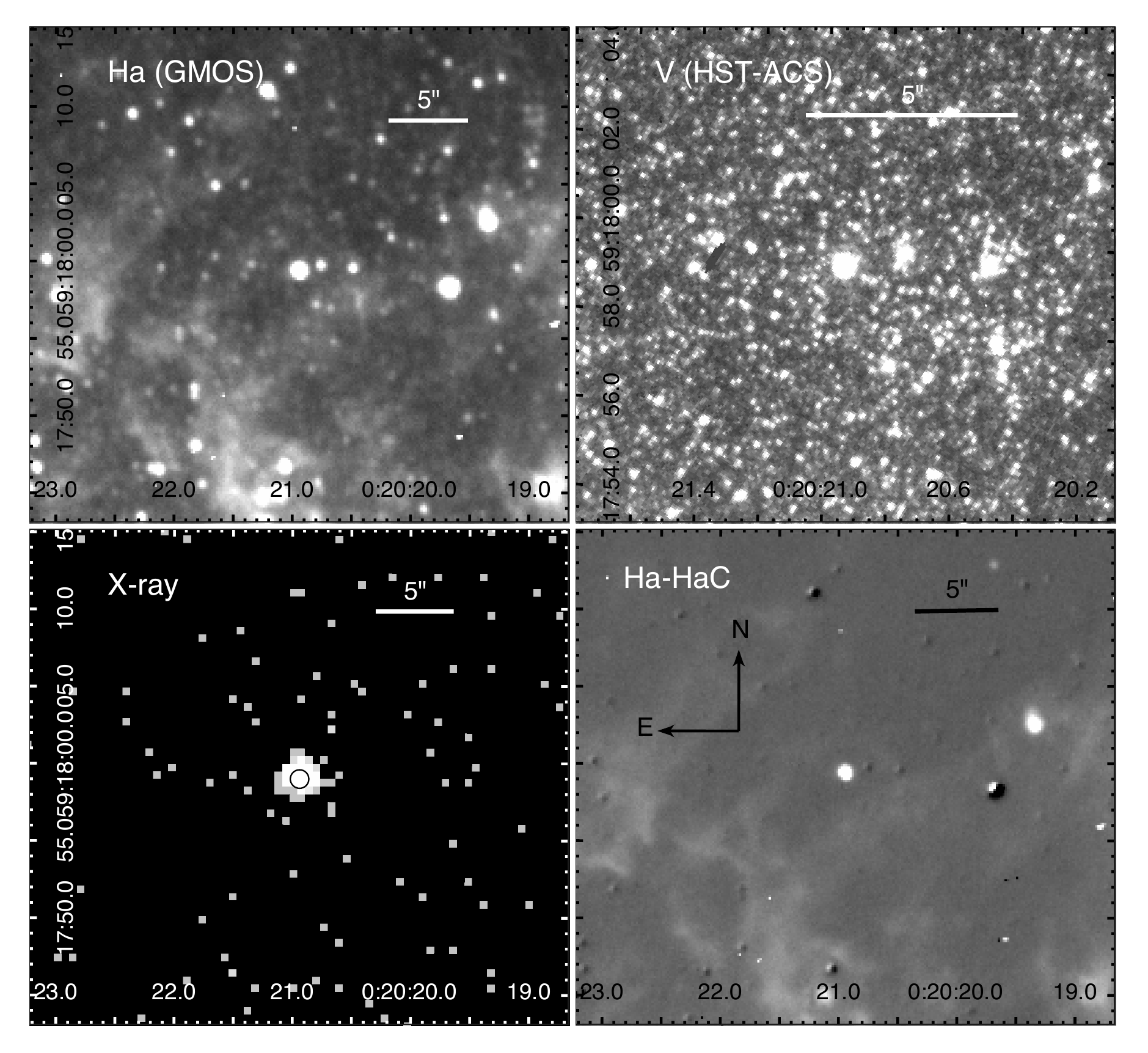}
\caption{{\bf  Optical and X-ray images of the field surrounding the transient X-ray source IC 10 X-2. Clockwise from upper left: Gemini H$\alpha$, HST-ACS V-band, Gemini H$\alpha$-H$\alpha$C difference-image, Chandra ACIS broad-band image.  The circle on the lower left image denotes the 95\% confidence region. All images share the same scale and orientation except the HST-V image which is magnified (see the 5'' scale bars in each image).  }}
\label{fig:Images}
\end{center}
\end{figure*}

Narrow band H$\alpha$ and H$\alpha_{C}$ Gemini Multi Object Spectrograph (GMOS) images from the Gemini Science Archive (GSA)\footnote{http://www3.cadc-ccda.hia-iha.nrc-cnrc.gc.ca/en/gsa} show the counterpart as a strong H$\alpha$ emitter. Figure \ref{fig:Images} shows the direct and difference images, with the Chandra error circle for comparison. The GMOS images were astrometrically calibrated using the catalog of \cite{massey2007a} and differenced in IRAF 2.14\footnote{http://iraf.noao.edu}. The Gemini  H$\alpha_{C}$ filter is  designed to isolate an equal region of continuum emission immediately adjacent to the H$\alpha$ line. The on and off-line images were obtained minutes apart, with the same exposure time (1800s), under excellent seeing (FWHM=0.5''). In the difference image H$\alpha$ line emission appears as positive (white) features, while absorption is negative (black). A single prominent emission feature is seen at the coordinates of the star identified above; its profile is consistent with a point source.
Additional images from the Hubble Space Telescope (2002), Mayall 4-m at KPNO (2000), and Gemini (2006-9)  were examined to check for proper motion (which would indicate a foreground object) and variability.

% Figure 2 

We used GMOS to obtain a spectrum of this optical counterpart (as part of a 25-object mask).  Our MOS mask setup was observed 6 times over two different nights (2010, September 2nd and 7th).  Each exposure was 45 min, for a total integration time of 4 hours in dark, photometric conditions with good seeing (CC50/IQ70/SB20)\footnote{see the Gemini or GSA webpages for observing condition definitions}. We used the B600 grating centered at 6000$\AA$ and 6050$\AA$ with 0.75" slitlets, and the detector binned 2$\times$2 to provide spectral resolution of 0.9$\AA$/pix (equivalent to 45 km/s). The central wavelength was carefully chosen so as to observe the H$\alpha$ line in all 25 objects, while reaching blue-ward of H$\beta$ for as many stars as possible.  Two closely spaced wavelength settings were used in order to cover the CCD gaps in the GMOS focal plane. Due to the properties of GMOS it is not possible to obtain the same wavelength coverage for objects in different parts the focal plane. 

GMOS spectra were reduced using the Gemini-IRAF package. Processing consisted of: flat-field correction, cosmic-ray rejection, wavelength calibration using GCAL CuAr arc exposures, rectification to a linear wavelength vs spatial coordinate system, 2-D sky subtraction, and extraction to 1-D spectra. We analyzed the spectrum of each object using IRAF {\it splot} and {\em R}.

\section{Results}
\subsection{The X-ray Properties of   IC 10 X-2, (=CXOU J002020.99+591758.6, =W05[XA-11] )  }
\label{xrayresults}
%% Object 46 in our Chandra catalog

The aspect corrected Chandra coordinates for IC 10 X-2 are RA: 00$^h$20$^m$20$^s$.94, Dec: 59$^\circ$17$^m$59$^s$.0, with a 95\% confidence region of radius (statistical plus systematic) $\sqrt{0.31^2 + 0.5^2}$ =  0.6''. The object was detected in outburst during the 2003 Chandra ACIS-S observation (30 ksec integration) and appears as object XA-11 in \cite{wang2005} and as IC 10 X-2 in \cite{Liu2011} however its true nature was not recognizable from a single observation. The long-term light curve is presented  in Figure~\ref{fig:monitor}, showing that we detected X-2 on three occasions, and placed upper limits on its quiescent luminosity on a further 7 occasions.

\begin{figure}
\begin{center}
\includegraphics[angle=0,width=8.5cm]{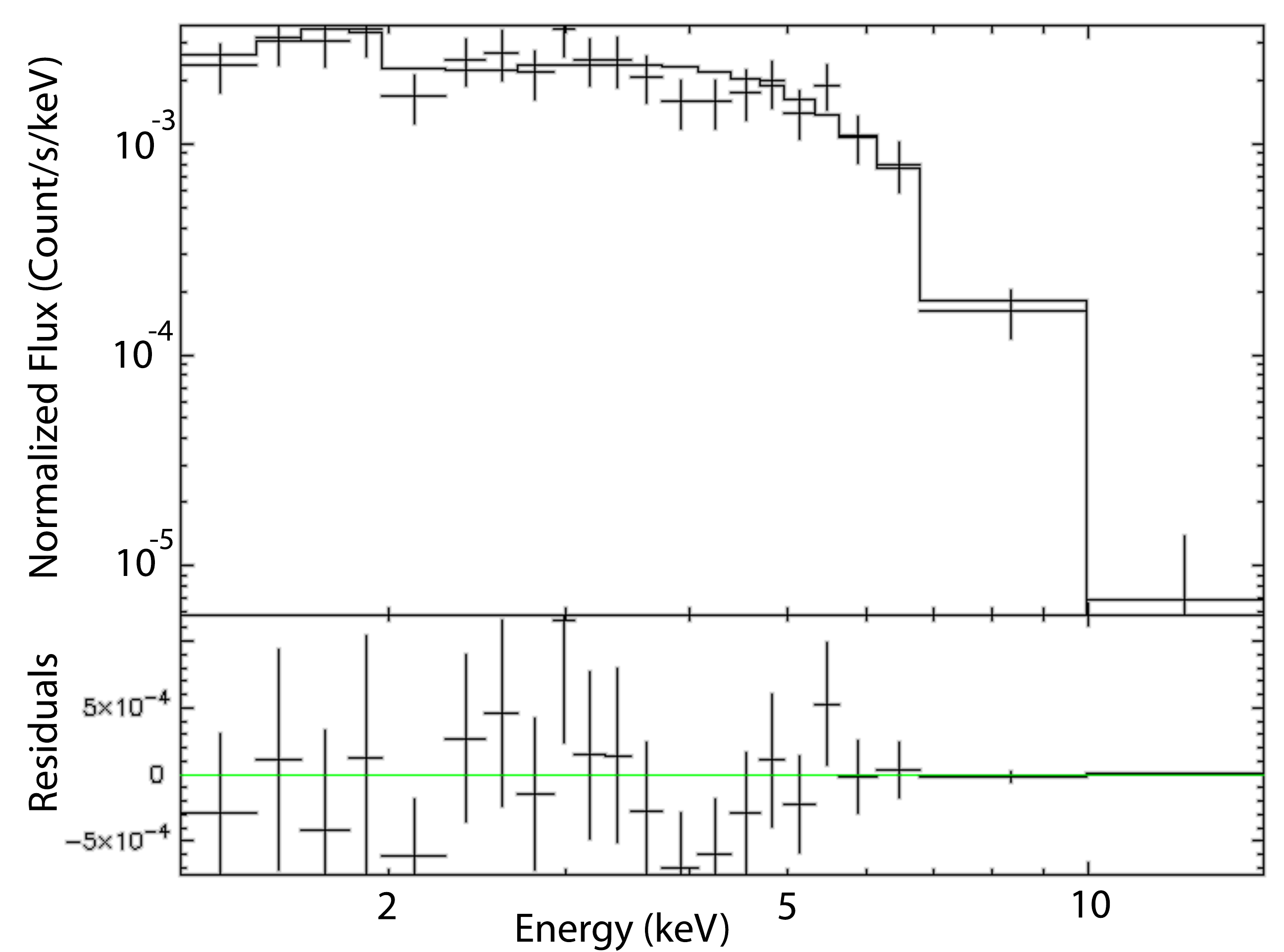}
\caption{{\bf  Chandra ACIS-S X-ray Spectrum. Extracted from the 2003 exposure (30 ksec integration) when the source was in outburst at $>$10$^{37}$\lx.  The best fitting absorbed powerlaw spectral model is shown.}}
\label{fig:xspec}
\end{center}
\end{figure}

During the initial 2003 Chandra observation, X-2 was the second brightest X-ray object in IC 10, after the massive black hole binary X1 \citep{Prestwich2007}. The net broad-band count rate of 1.18 $\times$10$^{-2}$ count s$^{-1}$ corresponds to an X-ray luminosity of 1.8$\times$10$^{37}$\lx  for the best fitting power-law spectral model ($\Gamma$=0.3, $N_H$=6.3$\times$10$^{21}$ atom cm$^{-2}$) shown in Figure~\ref{fig:xspec}.  We tested several spectral models using XSPEC  as summarized in Table~\ref{tab:xspec}. These were a simple power-law appropriate for X-ray binaries, a blackbody and thermal bremsstrahlung models (both widely applicable to stellar coronae, LMXBs and BH systems), and NEI plasma (non equilibrium ionization) which is descriptive of supernova remnants. All fits included the {\it phabs} component for foreground absorption. All models were able to reasonably represent the data (reduced $\chi^2 \sim$1), but examination of the parameter values definitively rules out the TB and NEI models, which required unrealistically high temperatures (kT$\gtrsim$100 keV).

% Figure 3  

%%%% TABLE 2 %%%%%%%%%%
\begin{deluxetable}{lllll}
\tablecaption{X-ray Spectral Fits \label{tab:xspec}} 
%\tablewidth{0pt}
\tablehead{
\colhead{Model}	      & \colhead{Parameters}   	          	   & \colhead{$\chi^2$ / DoF}                                        &  \colhead{$f_X$ (2-10 keV)}\\
\colhead{}                       &     \colhead{\tablenotemark{a}}             &     \colhead{\tablenotemark{b}}                               &      \colhead{erg cm$^{-2}$s$^{-1}$}           \\
 }
\startdata 
Powerlaw                    &  $\Gamma$ =  0.3           			&  			21/20		& 3.5$\times$10$^{-13}$ \\ 
				  & $N_H$ = 6.3$\times$10$^{21}$  		&					&  					\\
\tableline 
Blackbody       		&	$k$T	 = 2.2 keV					&		17/20		&		3.1$\times$10$^{-13}$		\\	
				&	$N_H$ = 2$\times$10$^{21}$					&					&				\\
\tableline 
Thermal Brems.       &	$k$T	 = 200 keV			   &		27/20			&		2.3$\times$10$^{-13}$		\\	
				&	$N_H$ = 1.6$\times$10$^{22}$ &		&				\\
\tableline 
NEI Plasma       	&	$k$T	 = 80 keV					&		24/19			&			2.4$\times$10$^{-13}$	\\
				&	$\tau$ = 5$\times$10$^{11}$					&					&				\\	
				&	$N_H$ = 1.6 $\times$10$^{22}$					&					&				\\								
\enddata
\tablenotetext{a}{ Values of spectral model parameters rounded to appropriate significant figures.}
\tablenotetext{b}{ Fits are for 228 net counts, 0.3-8 keV, grouped in 23 bins}

\end{deluxetable}

By Nov 2006 the source was barely detectable in the pair of 45 ksec observations (separated by 2 days). It was detected by wavdetect only in the first of the two pointings, in the hard band only. Stacking the two exposures to 90 ksec allowed a 3.8$\sigma$ flux measurement in the broad-band of 1.65$\times$10$^{-4}$ count s$^{-1}$.  Subsequent observations were shallower (15 ksec integrations) and provide 6 null-detections (none of which is more constraining than than 3$\times$10$^{-4}$ count s$^{-1}$) and one positive detection at 2.16$\times$10$^{-3}$  count s$^{-1}$. Taking the ratio of the peak to the lowest B-band count-rate, provides a variability factor of $f_{max}$/$f_{min}$ = 71. We note that if we did not stack the data, then our conservative (5 counts) upper limit estimate for non-detection in 45 ksec would impose an upper limit of $f_{max}$/$f_{ulim}$ = 117. Thus the result for either method agrees within a factor of $<$2. Applying the same spectral model as observed  at peak brightness, to the faint 2006 detection gives a luminosity of 1.8$\times$10$^{35}$\lx. This could represent the quiescent luminosity of a faint HMXB transient, alternatively it could be that the source was in transition to or from an active state. 
From the 2009-2010 monitoring series, we can only constrain the overall duration of the May 2010 outburst  (MJD 55337.5) to be less than 107 days. The source was not yet detectable on MJD 55290 and had faded from view by MJD 55397.

The median photon energies for the three positive detections, are 3.3$\pm$0.1 keV, 6.2$\pm$0.5 keV, and 3.4$\pm$0.3 keV, respectively. A K-S test confirms that the photon energy distribution during the May 2010 detection is consistent with the 2003 outburst (p=0.58), and supports the faint 2003 spectrum being harder (p=0.0012); despite the small number of counts.

% Figure 4   
\begin{figure}
\begin{center}
\includegraphics[angle=0,width=8.5cm]{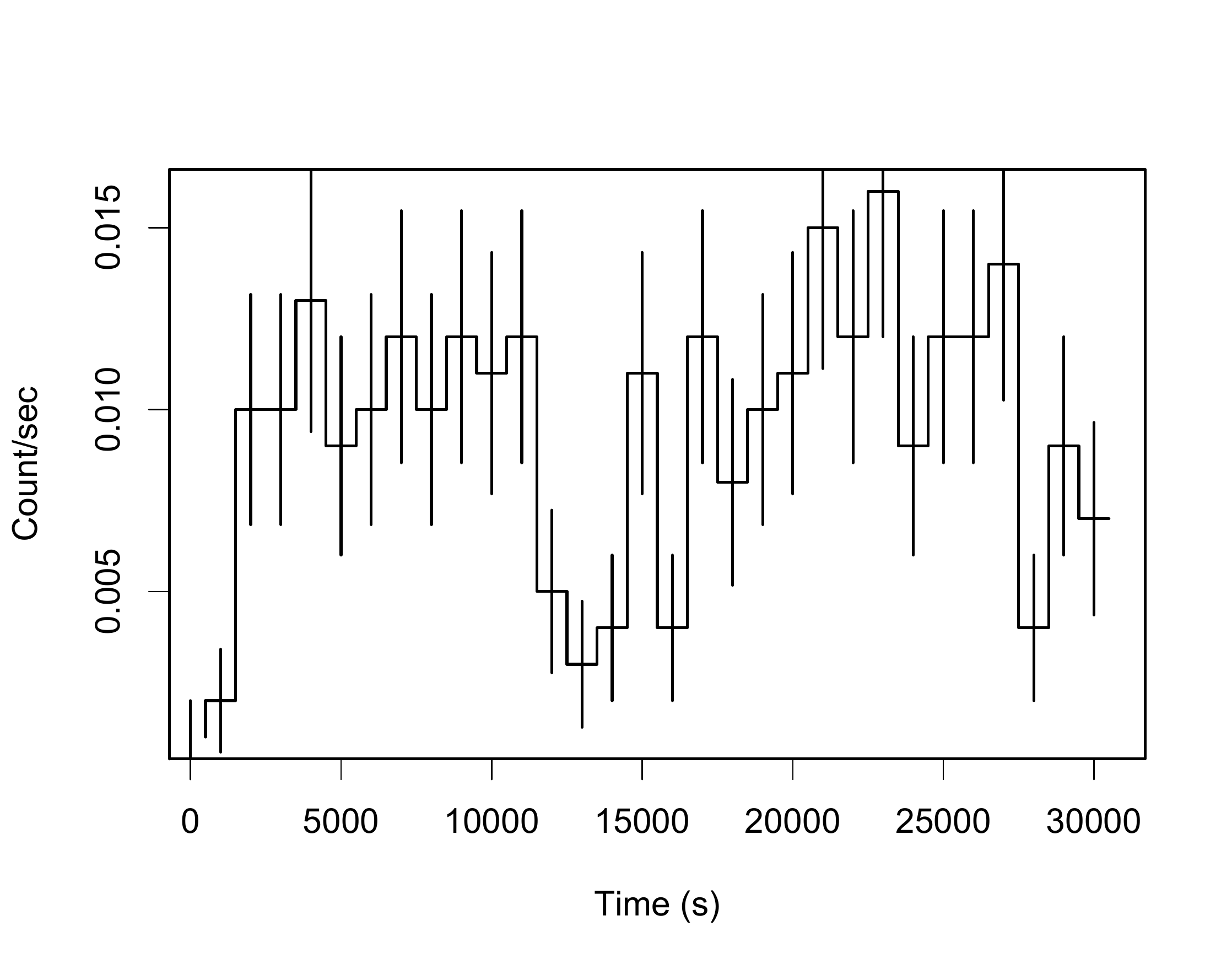}
\caption{{\bf Broad-band (0.3-8.0 keV) X-ray Lightcurve during the 30 ksec Chandra ACIS-S observation of 2003. The events are binned to 1 ksec resolution, with poisson error bars.}}  
\label{fig:binned_lc}
\end{center}
\end{figure}

For the bright outburst of 2003, the broad-band (0.3-8.0 keV)  lightcurve is shown in Figure~\ref{fig:binned_lc}. Events are binned to 1000 sec resolution, giving an average of $\sim$10 events per bin. The lightcurve reveals variability at $>$ 3$\sigma$ level(assuming $\sqrt{N}$ poisson errors).   A K-S test performed on 326 ACIS event arrival times, returns D=0.068, p=0.1 compared to a uniform distribution of arrival times. Thus variability is likely present, but we refrain from reading too much into the shape of the lightcurve. We searched for periodicity using the Lomb-Scargle periodogram (range = 6.5s-10,000s), no features rise to the 90\% significance level for a blind search using the method of \citep{NumRec}.

\subsection{Optical Counterpart: [LGGS]J002020.94+591759.3,  [2MASS] J002020.90+591759.0	}
IC 10 X-2 is positionally coincident with the star [LGGS]J002020.94+591759.3  in the catalog of \cite{massey2007a}. The distance between the X-ray and optical coordinates is 0.28'', which is within the 95\% error circle. There is a 2MASS point source catalog \citep{2mass} counterpart [2MASS] J002020.90+591759.0 which lies 0.31" away from our X-ray position. The association with this 2MASS star was noted by \cite{wang2005}, and our result moves the X-ray coordinates closer toward it. The star appears in the \cite{massey2007b} catalog of emission line objects, based upon narrow-band photometry, it does not appear in the \cite{crowther2003} catalog of WR stars in IC 10. 

% Figure 5  
\begin{figure}
\begin{center}
\includegraphics[angle=0,width=9cm]{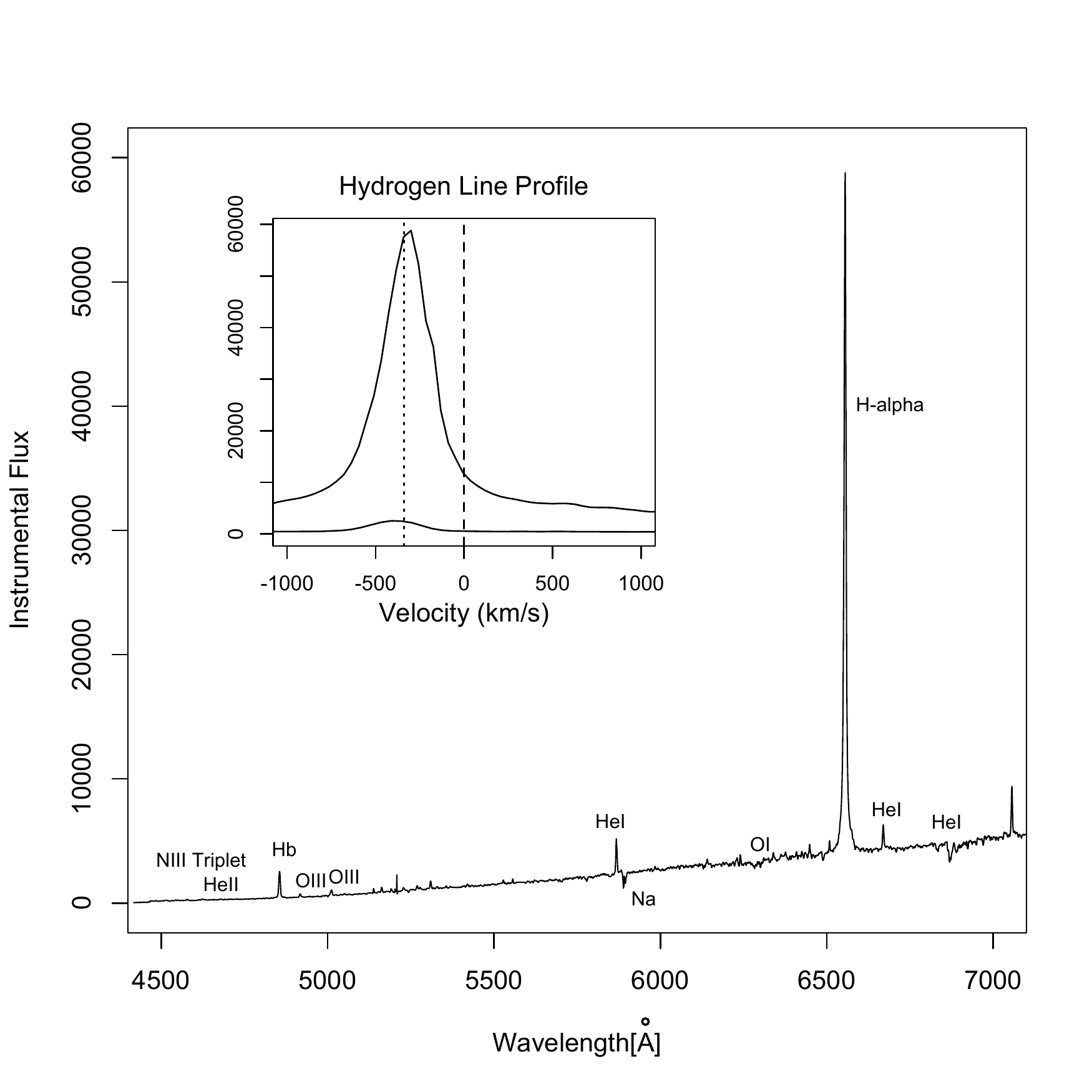}
\caption{{\bf  Gemini GMOS optical spectrum. Strong Balmer emission lines are seen, accompanied by numerous weaker lines of HeI, HeII, OIII, NIII, etc. The continuum is smooth and contains no strong metal lines.  Inset shows the velocity profiles of the H$\alpha$ and H$\beta$ lines, which are blue-shifted by the -340 km/s radial velocity of their host galaxy (IC 10). The spectrum is not flux calibrated, the Instrumental flux units are ADU per binned detector pixel.}}
\label{fig:gmos_spec}
\end{center}
\end{figure}

The optical spectrum (Fig~\ref{fig:gmos_spec}) resembles a Be star, with its prominent hydrogen-Balmer emission lines, and smooth continuum. The average blueshift of the Balmer lines gives a radial velocity of -337 km/s, which places the object in IC 10.  In Table~\ref{tab:phot} we show the observed and de-reddened colors using extinction values from the literature, and derived by us from the X-ray spectrum. The apparent magnitude is V=19.95, which applying $\mu$=24.1, $A_V$=3.1 results in absolute magnitude $M_V$= -7.2. This is in the upper range of luminosity for supergiants. The de-reddened optical colors are too red for a normal OB spectral type \citep{allen}, with the exception of the U-B color, which indicates a hot star. We speculate that the colors could be affected by circumstellar material, and the large equivalent width of H$\alpha$ which contributes significantly to the R-band luminosity.  

%%%%%%% TABLE 3 %%%%%   
\begin{deluxetable}{lllll}
\tablecaption{Optical and IR Magnitudes\label{tab:phot}} 
%\tablewidth{0pt}
\tablehead{
\colhead{Parameter} & \colhead{Observed}   & \colhead{$A_{opt}$} & \colhead{$A_X$} \\
\colhead{} &                   \colhead{\tablenotemark{a}} &                      \colhead{\tablenotemark{b}} &                   \colhead{\tablenotemark{c}} 
 }
\startdata     
$M_V$      & 				  & -6.78	& -7.51  \\
V                &  19.954$\pm$0.005 & 17.32             &   16.59      \\
 U-B            & -0.259$\pm$0.011 & -0.87   &  -1.04       \\
B-V              &1.211$\pm$0.005   & 0.32   &  0.07       \\ 
V-R              & 1.043$\pm$0.005  & 0.39   &  0.20       \\
R-I                & 0.820$\pm$ 0.005  & 0.10  &  -0.095      \\
V-K		& 4.454                           & 2.11  & 1.47 	\\
J-H		& 1.0			             &	0.76 & 0.69      \\
J-K             & 0.9                                &  0.45  & 0.33  \\
H$\alpha$-R & -1.16                        &    \tablenotemark{d}       &\tablenotemark{d}
\enddata
 \tablenotetext{a}{ References: UBVRI: \cite{massey2007a}, JHK: 2MASS, H$\alpha$: \cite{massey2007b} }
 \tablenotetext{b}{De-reddened values using:  optically derived E(B-V)=0.85 \citep{Sanna2008} }
 \tablenotetext{c}{X-ray derived E(B-V)=1.09 (this study)  }
 \tablenotetext{d}{No straightforward correction of H$\alpha$-R for extinction.}

\end{deluxetable}

The Balmer lines are strongly in emission with EW[H$\alpha$, $\beta$] = -110 \AA, -25\AA. At the blue end of our GMOS spectral range, we detect the combination of  NIII (4634-4640-4650) and HeII (4686) emission lines that constitute the primary characteristic of spectral class Of. According to \cite{gray2009} the Of stars are regarded as a supergiant analog of Be stars, having significant outflows. 

\begin{figure}
\begin{center}
\includegraphics[angle=0,width=9cm]{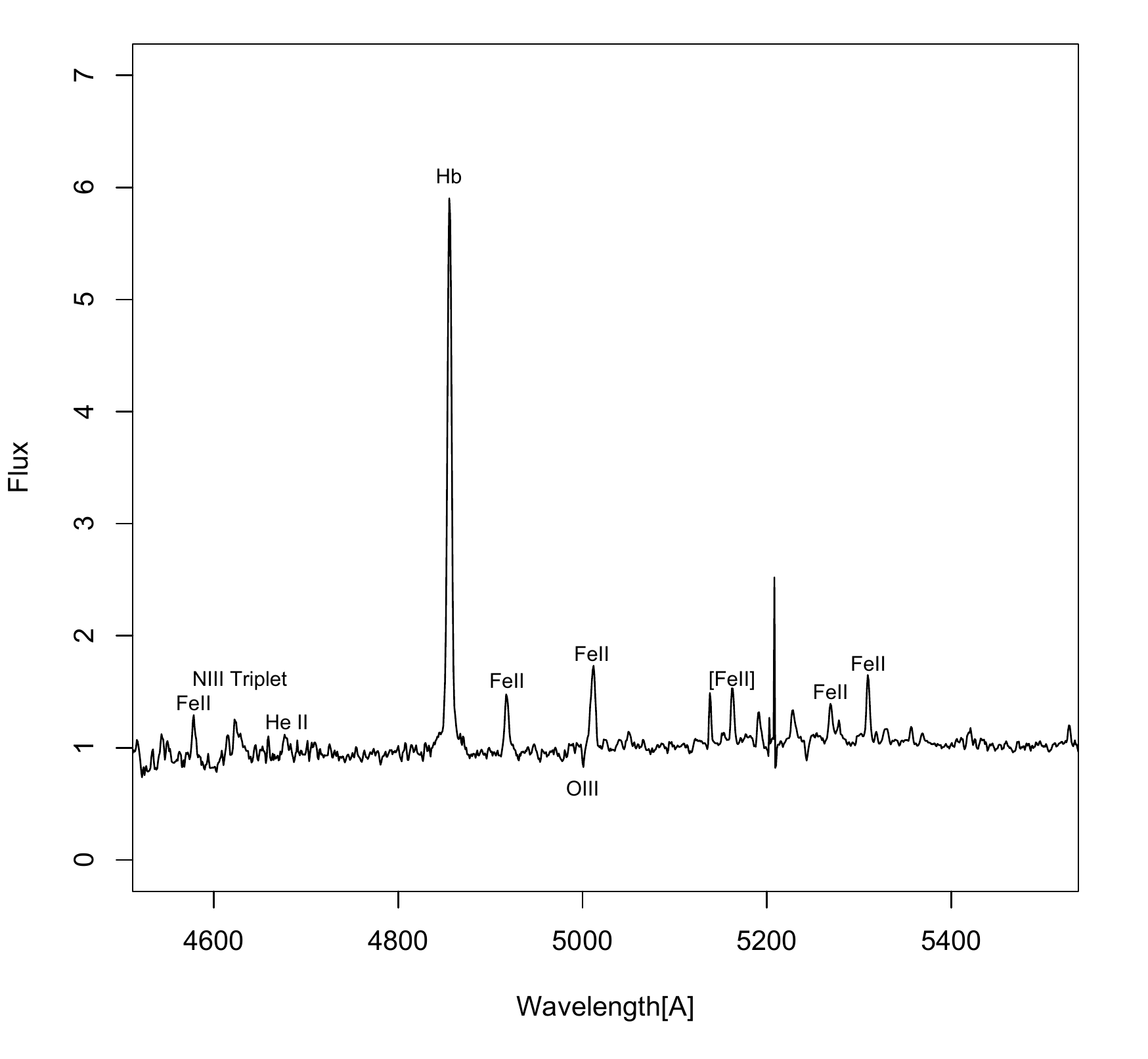}
\caption{{\bf  Optical spectrum in the 4000-5000  region, showing the combination of NIII ($\lambda \lambda$4634-4640-4642 \AA) and HeII ($\lambda\lambda$4686 \AA) emission lines, that constitute the primary characteristic of spectral class Of, and in addition, the complex of Fe II permitted and forbidden lines characteristic of spectral class B[e].}}
\label{fig:gmos_spec_blue}
\end{center}
\end{figure}

The star also shows the complex of Fe lines that characterize the B[e] spectral class. In Figure~\ref{fig:gmos_spec_blue} we identify both permitted and forbidden lines of Fe. The H$\alpha$ and H$\beta$ lines both show broad wings indicative of high velocity bulk flows, but there is no double peak that would indicate an inclined circumstellar disk. 

The star shows selective emission and absorption effects in the HeI lines; with HeI ($\lambda\lambda$ 5865) in emission and HeI ($\lambda\lambda$ 6875) in absorption (See Fig~\ref{fig:gmos_spec}). The physical conditions leading to the appearance of emission lines from certain energy levels, while other transitions in the same ion species remain in absorption is not well understood. However the phenomenon is a well established feature of the O-type stars (Walborn 2001), and the selective-emission lines are believed to originate in the stellar photosphere.

% Figure 6 

According to the classification system of  \cite{walborn1976}, supergiant O and B stars are divided into the OBN and OBC subtypes. By analogy to the classification protocol for WR stars, the OB-N/C subtypes denote nitrogen or carbon enrichment.  Stars of this class show He lines, but usually in absorption, except at the highest luminosities. The identification of the NIII triplet puts the star in the OBN class of nitrogen enriched B supergiants. According to \cite{Walborn1977} the ``f" subclass corresponds to N IV 4058 stronger than the NIII 4634-460-4642  triplet; our spectrum does not reach shortward of 4500\AA. The current practice forgoes the OBC designation as merely a normal OB star, but recognizes the unusual abundance pattern of OBN stars. In this scheme (E.g. \citealt{massey2009, gray2009}) the combination of NIII and HeII 4686 in emission identify the ``f" subclass.

 The first star later than O8 to show He II 4686 emission was HD 226868  (the primary in the black hole HMXB Cyg X-1). Selective emission phenomena are functions of temperature and luminosity with a distinct diagonal boundary in a 2D plane \citep{walborn1971}; such that this feature at a relatively low temperature (a late O or early B spectral type), implies a superluminous object. Figure 5 of \cite{Walborn1977} compares the absolute visual magnitude of LMC and Galactic 09-B0 supergiants, which range from -6.0 (II) to -8.0(Ia+). So this star at $M_V$=-7.2 is in luminosity class Ia, which is consistent with the appearance of  He II 4686 in emission. 

\cite{massey2009} finds that lower metallicity (in the SMC) results in stars that are over-luminous (by a whole luminosity class or more) compared to what their spectral classifications would imply. This is consistent with the observation of \cite{Walborn1977}, that supergiants in the LMC are more luminous than those of the same spectral type in the MW.  Lower metallicity results in weaker stellar winds, thus for a metal-poor star to exhibit the ``usual" luminosity indicators requires a much larger luminosity than if it had ``normal" (galactic) metallicity.  For example the SMC star AzV 83 has both N III 4638-42 and He II 4686 strongly in emission, leading to an OIf or even OIaf luminosity class, given the strength of emission \citep{massey2009}.  This argument would apply to a lesser extent in IC10, as the metallicity is intermediate between SMC and the Milky Way.

In the \cite{Walborn2000} atlas of peculiar OB spectra, the WN stars appear similar to our object in displaying HIII, HeII and H$\beta$ emission. Unfortunately they don't go longward of H$\beta$ so we cannot compare the complex of Fe lines. 

\section{Discussion and Conclusions}

The bright X-ray transient IC 10 X-2 is located close to the center of IC 10. A positional optical counterpart lies within the 0.31" error radius. This counterpart shows prominent Balmer lines in emission, doppler-shifted to a radial velocity of  -340 km/s matching the established value for IC 10. Analysis of the X-ray light curve, reveals a factor $\sim$100 variability, and a hard power law spectrum, with absorption consistent with IC 10. The peak X-ray luminosity implied by $D_{IC10}$=660 kpc is 1.8$\times$10$^{37}$\lx.  The apparent magnitude, distance and reddening imply a supergiant primary with absolute magnitude in the range $M_V$=-6.8 \textbf{to} -7.5 depending on the extinction model used. The optical spectrum is dominated by Balmer emission lines, a faint continuum is seen, and the photospheric emission lines NIII[4634-4640-4642] triplet, HeII[4686] are present, which are indicative of spectral type Of I-II. The star may be over luminous for its spectral type due to the low metallicity of IC 10. 

The very large equivalent width of the optical emission lines, and the IR excess provides abundant evidence for an enhanced stellar outflow, as is typically seen in HMXB primaries (e.g. \citealt{wilson2008}). Thus the star's mass-loss is sufficient to provide the required accretion rate onto the NS. In order for the circumstellar material to form strong optical emission lines without raising the X-ray absorption above 10$^{22}$ atom cm$^2$, it must lie out of the line of sight to the X-ray source. A circumstellar disk could provide the required geometry if it were oriented close to face-on. Galactic OBN stars are preferentially found in close binaries, where the enrichment of nitrogen is due to tidal dredge-up of CNO cycle products from the core.  Rotational mixing is now recognized as a key factor in the formation of massive binaries \citep{deMink2013}. According to  \cite{Bolton1978} between 50\% and 100\% of the OBN stars appear to be short-period binaries (P$\sim$1-100d) while none of the OBC stars fall into this category.  The counterpart to X-2 fits this pattern, as the measured X-ray luminosity requires an XRB interpretation, which in turn implies a close binary. The binary must be close in order for the wind density \citep{Chlebowski1991} encountered by the compact object to deliver a sufficient mass accretion rate.  The range of binary separations for supergiant HMXBs in the Milky way is 0.04-0.8 A.U. (typically a few $R_*$), corresponding orbital periods  are 1-200 days (See Table in \cite{Lewin1995}).

Our formation scenario for this system, is that the precursor was a close OBN+O star binary, whose orbit was both widened and made elliptical by the supernova kick. This could be a similar system to HD235679, a supergiant Be star in a spectroscopic binary (P=225 d) with an invisible companion \citep{Bolton2001}.  Difficulty remains in in reconciling the transient X-ray outbursts with a powerful wind, which must be supplying a large mass transfer rate at all times. SG-HMXBs are usually continuous wind accretors; the same paradox applies to SFXTs \citep{romano2013}. Plausible solutions include a highly eccentric orbit to modulate the mass transfer rate, as in Be-HMXBs, or a rapidly spinning young neutron star that does not accrete because  is evolving through the so-called propellor regime. Supergiant+NS binary systems where the primary overflows its Roche-lobe, yet accretion is centrifugally inhibited are explicitly predicted by the theoretical work of \cite{urpin1998}. The duration of this ``disk-propellor" phase is expected to be $\sim$10$^4$ yrs, followed by onset of a ``standard" SG-HMXB phase if the NS spin is spun-down sufficiently.  Assuming the primary to be a 30$M_{\sun}$ supergiant with $M_V$=-7.5, and applying the standard approximation for main sequence lifetime $T_{ms} \sim \frac{M}{L} \times$10$^{10}$ yr, gives an age of 3.5My. Thus it is possible that in IC 10 X-2 a very young NS is  spinning at a rate that puts it in the centrifugal inhibition regime. If true then the spin period is probably $<$1s and the orbital period should be a few days at most.

Alternative explanations for IC10 X-2 include that the optical counterpart is not a single star, but an association buried in a compact HII region. A  Be-NS HMXB could then explain the transient X-ray emission, although the X-ray spectrum would then be much harder than is usual for such systems. The supernova imposter SN2010da  \citep{Binder2011} is similar in this regard.  Attempts to explain the large equivalent widths as a SNR, PNe or HII  region are inconsistent with the absolute magnitude of -7, and compact dimensions. Images from the {\it HST} ACS show an optical point-source with no extended nebulosity. At 660 kpc, the projected size of the HST FWHM ($\sim$0.1 arc sec) is 0.32 pc. This is  smaller than typical HII regions \citep{HongLee1990}. A foreground galactic star is ruled out by the RV value.

\section{Acknowledgements} 
We thank A. Camero, B. F. Williams , and A. Prestwich for useful discussions. 
We acknowledge the support of SAO grant NASA-03060 from the Chandra X-ray Observatory.
SL thanks Gemini Observatory for supporting this research. Gemini is operated by the Association of Universities 
for Research in Astronomy, Inc., on behalf of the international Gemini partnership of Argentina, 
Australia, Brazil, Canada, Chile, the United Kingdom, and the United States of America.


\begin{thebibliography}{}

\bibitem[Allen(2000)]{allen}
Allen's Astrophysical Quantities, 4th edition, Cox, A., (Editor), Springer-Verlag, New York. 

\bibitem[Antoniou et al.(2010)]{antoniou2010} 
Antoniou, V., Zezas, A., Hatzidimitriou, D., \& Kalogera, V.\ 2010, \apjl, 716, L140 

%%XSPEC: The First Ten Years
%\bibitem[Arnaud(1996)]{xspec} 
%Arnaud, K.~A.\ 1996, Astronomical Data Analysis Software and Systems V, 101, 17 

% HEAD MEETING 2011, poster
%\bibitem[Balchunas et al.(2011)]{Balchunas2011}
%Balchunas, A., Laycock, S., ... , 2011, HEAD 

\bibitem[Bauer \& Brandt(2004)]{bauerbrandt2004}
Bauer, F. E., \& Brandt, W. N., 2004, \apj, 601, L67

%On the Apparent Lack of Be X-Ray Binaries with Black Holes
\bibitem[Belczynski \& Ziolkowski(2009)]{belczynski2009} 
Belczynski, K., \& Ziolkowski, J.\ 2009, \apj, 707, 870 

%CHANDRA DETECTION OF SN 2010da FOUR MONTHS AFTER OUTBURST: EVIDENCE FOR A HIGH-MASS X-RAY BINARY IN NGC 300
\bibitem[Binder et al.(2011)]{Binder2011} 
Binder, B., Williams, B.~F., Kong, A. K., Gaetz, T. J., Plucinsky, P. P., Dalcanton, J. J., Weisz, D. R., \ 2011, \apj, 739, L51 

% Supergiant Be star spectroscopic binary with invisible companion
\bibitem[Bolton \& Hurkens(2001)]{Bolton2001} 
Bolton, C.~T., \& Hurkens, R.\ 2001, Publications of the Astronomical Institute of the Czechoslovak Academy of Sciences, 89, 23 

%% The binary frequency of the OBN and OBC stars
\bibitem[Bolton \& Rogers(1978)]{Bolton1978} 
Bolton, C.~T., \& Rogers, G.~L.\ 1978, \apj, 222, 234 

% Extinction Law
%\bibitem[Cardelli, Clayton \& Mathis]{cardelli1989}
%Cardelli, Clayton \& Mathis, 1989, \apj, 345, 245


%% On winds and X-rays of O-type stars
\bibitem[Chlebowski \& Garmany(1991)]{Chlebowski1991} 
Chlebowski, T., \& Garmany, C.~D.\ 1991, \apj, 368, 241
   
%On the Wolf-Rayet counterpart to IC 10 X-1
%\bibitem[Clark \& Crowther(2004)]{clark2004} 
%Clark, J.~S., \& Crowther, P.~A.\ 2004, \aap, 414, L45    

%\bibitem[Corbet(1984)]{corbet1984}
%Corbet, R. H. D., 1984, A \& A, 141, 91
   
%Gemini observations of Wolf-Rayet stars in the Local Group starburst galaxy IC 10   
\bibitem[Crowther et al.(2003)]{crowther2003} 
Crowther, P.~A., Drissen, L., Abbott, J.~B., Royer, P., \& Smartt, S.~J.\ 2003, \aap, 404, 483 
   
%The Rotation Rates of Massive Stars: The Role of Binary Interaction through Tides, Mass Transfer, and Mergers
\bibitem[de Mink et al.(2013)]{deMink2013} 
de Mink, S.~E., Langer, N., Izzard, R.~G., Sana, H., \& de Koter, A.\ 2013, \apj, 764, 166 

 %Discovery in IC10 of the farthest known symbiotic star (Good reference for Extinction to IC 10)
\bibitem[Gon{\c c}alves et al.(2008)]{Goncalves2008} 
Gon{\c c}alves, D.~R., Magrini, L., Munari, U., Corradi, R.~L.~M., \& Costa, R.~D.~D.\ 2008, \mnras, 391, L84    
  
%Discovering a 5.72-d period in the supergiant fast X-ray transient AX J1845.0-0433  
\bibitem[Goossens et al.(2013)]{Goosens2013} 
Goossens, M.~E., Bird,  A.~J., Drave, S.~P., et al.\ 2013, \mnras, 434, 2182     
      
\bibitem[Gray(2009)]{gray2009}   
Gray, R. O., 2009, ``Stellar Spectral Classification"  
   
%High-mass X-ray binaries as a star formation rate indicator in distant galaxies   
\bibitem[Grimm et al.(2003)]{GGS} 
Grimm, H.-J., Gilfanov, M., \& Sunyaev, R.\ 2003, \mnras, 339, 793  
 
% The HII Regions of IC 10
\bibitem[Hodge \& Lee(1990)]{HongLee1990} 
Hodge, P., \& Lee, M.~G.\ 1990, \pasp, 102, 26

%X-Ray Processing of ChaMPlane Fields: Methods and Initial Results for Selected Anti-Galactic Center Field
\bibitem[Hong et al.(2005)]{hong2005}
Hong, J., van den Berg, M., Schlegel, E. M., Grindlay, J. E., Koenig, X., Laycock, S., Zhao, P., 2005, \apj, 635, 907	

%Discovery of a short orbital period in the Supergiant Fast X-ray Transient IGR J16479-4514
\bibitem[Jain et al.(2009)]{Jain2009} 
Jain, C., Paul, B., \& Dutta, A.\ 2009, \mnras, 397, L11

% AAS Winter Meeting 2010, poster 
\bibitem[Laycock et al.(2010)]{Laycock2010}
Laycock, S., Camero, A., Wilson-Hodge, C.~A., et al.\ 2010, Bulletin of the American Astronomical Society, 42, \#419.04

%% X-ray Binaries Book
\bibitem[Lewin(1995)]{Lewin1995}
Lewin, W., in X-Ray Binaries, 1995, C.U.P., Lewin, van Paradijs, van den Heuvel, Eds. 

\bibitem[Liu et al.(2005)]{liu2005}
Liu, Q. Z., van Paradijs, J., \& van den Heuvel, E. P. J.: 2005, \aap,  442, 1135

%%Chandra ACIS Survey of X-ray Point Sources in 383 Nearby Galaxies. I. The Source Catalog
\bibitem[Liu(2011)]{Liu2011} 
Liu, J.\ 2011, \apjs, 192, 10

%%Composition of the Chandra ACIS contaminant (PS, 12 pages) 
% http://cxc.harvard.edu/ciao4.4/why/acisqecontam.html
%\bibitem[Marshall et al.(2003)]{ACIScontaminant}
%Marshall, H. L. , Tennant, A., Grant, C. E.,  Hitchcock, A. P., O'Dell, S.,  Plucinsky, P. P., 2003, astro-ph/0308332

%A Survey of Local Group Galaxies Currently Forming Stars. II. UBVRI Photometry of Stars in Seven Dwarfs and a Comparison of the Entire Sample
\bibitem[Massey et al.(2007a)]{massey2007a} 
Massey, P., Olsen, K.~A.~G., Hodge, P.~W., et al.\ 2007, \aj, 133, 2393 

%A Survey of Local Group Galaxies Currently Forming Stars. III. A Search for Luminous Blue Variables and Other H? Emission-Line Stars
\bibitem[Massey et al.(2007b)]{massey2007b} 
Massey, P., McNeill, R.~T., Olsen, K.~A.~G., et al.\ 2007, \aj, 134, 2474 

%The Physical Properties and Effective Temperature Scale of O-Type Stars as a Function of Metallicity. III. More Results From the Magellanic Clouds
\bibitem[Massey et al.(2009)]{massey2009} 
Massey, P., Zangari, A.~M., Morrell, N.~I., et al.\ 2009, \apj, 692, 618 

%\bibitem[McBride et al.(2007)]{mcbride2007}
%McBride, V. A., Coe, M. J., Bird, A. J., et al. 2007, MNRAS, 382, 743

% Spectral type distribution of SMC HMXBs
\bibitem[McBride et al.(2008)]{mcbride2008}
McBride, V. A., Coe, M. J., Negueruela, I., Schurch, M. P. E., McGowan, K. E., 2008,  \mnras, 388, 1198

% SFXTs
\bibitem[Negueruela et al.(2006)]{negueruela2005} 
Negueruela, I., Smith, D.~M., Reig, P., Chaty, S., \& Torrej{\'o}n, J.~M.\ 2006, The X-ray Universe 2005, 604, 165 

\bibitem[Okazaki \& Negueruela(2001)]{okazaki2001}
Okazaki, A., Negueruela, I., 2001, A\&A, 377, 161

% Numerical Recipies n Fortran77
\bibitem[Press et al.(1996)]{NumRec}
Press, W., H., Teukolsky, S., A., Vetterling, W,. T., Flannery, B., P., Numerical Recipes in Fortran, 2nd Edition, 1996, CUP.

%The Orbital Period of the Wolf-Rayet Binary IC 10 X-1: Dynamic Evidence that the Compact Object Is a Black Hole
\bibitem[Prestwich et al.(2007)]{Prestwich2007} 
Prestwich, A.~H., Kilgard, R., Crowther, P.~A., et al.\ 2007, \apjl, 669, L21 


%IGR J18483-0311: a new intermediate supergiant fast X-ray transient
%\bibitem[Rahoui \& Chaty(2008)]{rahoui2008} 
%Rahoui, F., \& Chaty, S.\ 2008, \aap, 492, 163 

\bibitem[Reig(2011)]{reig2011}
Reig, P., 2011, Astrophysics and Space Science, Volume 332, Issue 1, pp.1-29

%The 100-month Swift catalogue of supergiant fast X-ray transients.
\bibitem[Romano et al.(2014)]{Romano2014} 
Romano, P., Krimm, H.~A., Palmer, D.~M., et al.\ 2014, \aap, 562, A2 

% IGR J16479?4514
%\bibitem[Romano et al.(2009)]{Romano2009}
%Romano, P., Sidoli, L., Cusumano, G., et al. 2009, MNRAS, 399, 2021

% Disentangling the System Geometry of the Supergiant Fast X-Ray Transient IGR J11215-5952 with Swift
\bibitem[Romano et al.(2009)]{Romano2009} 
Romano, P., Sidoli, L., Cusumano, G., et al.\ 2009, \apj, 696, 2068 

% SFXTs Swift survey   
\bibitem[Romano et al.(2013)]{romano2013} 
Romano, P., Mangano, V., Ducci, L., et al.\ 2013, Advances in Space Research, 52, 1593 

%On the Distance and Reddening of the Starburst Galaxy IC 10
\bibitem[Sanna et al.(2008)]{Sanna2008} 
Sanna, N., Bono, G., Stetson, P.~B., et al.\ 2008, \apjl, 688, L69 


% Rotation in OB binary formation
%\bibitem[Sana et al.(2012)]{sanna2012} 
%Sana, H., de Mink, S.~E., de Koter, A., et al.\ 2012, Science, 337, 444 

% SFXTs
\bibitem[Sguera et al.(2005)]{sguera2005} 
Sguera, V., Barlow, E.~J., Bird, A.~J., et al.\ 2005, \aap, 444, 221 


% shortest SFXT period (no longer!)
%\bibitem[Sidoli et al.(2013)]{Sidoli2013} 
%Sidoli, L., Esposito, P., Sguera, V., et al.\ 2013, arXiv:1302.1702

%Monitoring Supergiant Fast X-Ray Transients with Swift. I. Behavior outside Outbursts
\bibitem[Sidoli et al.(2008)]{Sidoli2008} 
Sidoli, L., Romano, P., Mangano, V., et al.\ 2008, \apj, 687, 1230 

%Transient outburst mechanisms in Supergiant Fast X-ray Transients
\bibitem[Sidoli(2009)]{Sidoli2009} 
Sidoli, L.\ 2009, Advances in Space Research, 43, 1464 

% 2MASS
\bibitem[Skrutskie et al.(2006)]{2mass} 
Skrutskie, M.~F., Cutri, R.~M., Stiening, R., et al.\ 2006, \aj, 131, 1163 

% Evolution of Neutron Stars in HMXBs
\bibitem[Urpin et al.(1998)]{urpin1998}
Urpin, V., Konenkov, D., \& Geppert, U., 1998,  \mnras, 299, 73

%The OB Zoo: A Digital Atlas of Peculiar Spectra
\bibitem[Walborn \& Fitzpatrick(2000)]{Walborn2000} 
Walborn, N.~R., \& Fitzpatrick, E.~L.\ 2000, \pasp, 112, 50 


%Some Spectroscopic Characteristics of the OB Stars: an Investigation of the Space Distribution of Certain OB Stars and the Reference Frame of the Classification
\bibitem[Walborn(1971)]{walborn1971}
Walborn, N.~R., 1971, \apjs, 23, 257


% The OBC and OBN Stars
\bibitem[Walborn(1976)]{walborn1976}
Walborn, N.~R., 1976, \apj, 205, 419

%Spectral classification of O and B0 supergiants in the Magellanic Clouds
\bibitem[Walborn(1977)]{Walborn1977}
Walborn, N.~R.\ 1977, \apj, 215, 53

%IC 10 with Chandra and XMM
\bibitem[Wang et al.(2005)]{wang2005} 
Wang, Q.~D., Whitaker, K.~E., \& Williams, R.\ 2005, \mnras, 362, 1065 

% The Ages of High-mass X-Ray Binaries in NGC 2403 and NGC 300
\bibitem[Williams et al.(2013)]{williams2013} 
Williams, B.~F., Binder, B.~A., Dalcanton, J.~J., Eracleous, M., \& Dolphin, A.\ 2013, \apj, 772, 12 

%The Distance To IC 10 From Near-Infrared Observations of Cepheids
\bibitem[Wilson et al.(1996)]{wilson1996} 
Wilson, C.~D., Welch, D.~L., Reid, I.~N., Saha, A., \& Hoessel, J.\ 1996, \aj, 111, 1106 

\bibitem[Wilson et al.,(2008)]{wilson2008}
Wilson, C. A., Camero-Arranz, A., 2008, \apj, 678, 1263


\end{thebibliography}
\end{document}